# ALARM: Active LeArning of Rowhammer Mitigations

Amir Naseredini
University of Sussex
Royal Holloway University of London
sahnaseredini@gmail.com

Martin Berger
Montanarius Ltd
Turing Core, Huawei 2012 Labs, Huawei R&D UK
University of Sussex
contact@martinfriedrichberger.net

Matteo Sammartino
Royal Holloway University of London
University College London
matteo.sammartino@rhul.ac.uk

Shale Xiong
Arm Ltd
xiongshale@gmail.com

## ABSTRACT

Rowhammer is a serious security problem of contemporary dynamic random-access memory (DRAM) where reads or writes of bits can flip other bits. DRAM manufacturers add mitigations, but don't disclose details, making it difficult for customers to evaluate their efficacy. We present a tool, based on active learning, that automatically infers parameter of Rowhammer mitigations against synthetic models of modern DRAM.

## 1 INTRODUCTION

Rowhammer [18] is a security vulnerability that exploits physical effects in DRAM (= dynamic random-access memory). DRAM achieves high capacity by storing each individual bit with just a capacitor. Capacitors, however, are subject to interference by charge fluctuations caused by accessing nearby bits, so reading or writing a bit at one memory cell *flips* one or more bits at *distinct* neighbouring cells with non-negligible probability. This probability correlates not just with data stored in the neighbourhood, but also with frequency of reads and writes in the neighbourhood of a target bit (frequent access is called "hammering" the "victim") between two refreshes of the victim. Refreshing restores a memory cell's charge and dramatically reduces the probability of a bit flip for a short while. See [31] for details of the physics of (DRAM) insecurity, and [12] on how to execute a Rowhammer attack.

As of 2022, DRAM security is a cornerstone of computer security in all mainstream systems, because many mechanisms for enforcing software security, in particular page-tables and capabilities, are stored in DRAM. The DRAM attack vector remains wide open even for CHERI [33], the most rigorous, security-oriented computer architecture available on the market today [22]. In marked contrast, there is an optimistic view that "Rowhammer is solved"[1]. This belief is based on the idea that existing Rowhammer mitigations detect and prevent most attacks, and DRAM's ECC (= error correcting codes) 'mops up' bit flips that evade the mitigations. Given DRAM vendors' secrecy about mitigations, should we simply accept the optimistic view? There are reasons to be sceptical!

- The history of Rowhammer gives us no confidence, because so far, every generation of DRAM was successfully attacked [19], including DDR4, the last generation DRAM that is available on the market in appreciable quantities.
- The nature of semiconductor physics seems to suggest that interference will necessarily increase with higher silicon density. Indeed, modern DRAM is already so vulnerable that "involuntary hammering", for example from non-malicious cache coherence traffic, is a problem [21]. 3D stacking [23] of DRAM is likely to aggravate the problem.
- The nature of the DRAM market emphasises memory capacity over everything else, and all Rowhammer mitigations cost silicon area, hence reduce memory capacity. This incentivises manufacturers to spend transistors on memory capacity rather than robustness against Rowhammer.
- Mitigations can be powerful attack vectors [19].
- There is a widespread unease of "security by obscurity".

With the ever increasing dependency of modern society on computers, it is vital for DRAM users not to rely just on unverifiable DRAM manufacturer claims about invulnerability to Rowhammer. Instead, and following the folk wisdom to "trust, but verify", users need independent and principled tooling to understand and evaluate:

(1) The Rowhammer mitigation capabilities of a vendor's design.
(2) Rowhammer susceptibility of individual chips.

---
[1]Private communication with several well-known and influential industrial computer architects.

Given the complexity of that task, this means automated tooling. A similar sentiment is implicit in [10, 13]. The present work uses off-the-shelf machine-learning tools to automate (1).

*Rowhammer mitigations.* A widely held belief is that Rowhammer can be 'solved' with ECC DRAM. While this is true in the sense that we could throw enough redundancy at each bit to correct flips, it is false in practise: the redundancy to make this work consumes silicon area and hence reduces, often drastically so, memory capacity. An important insight of Shannon's is that ECC can be made more efficient against an explicit error model. Historically, in the context of DRAM, the term ECC memory has been used without an explicit error model. Reading between the lines, it is clear that the error model was always a non-malicious attacker, typically nature (e.g., background radiation or cosmic particles) which flips bits IID (= independent and identically distributed). ECC efficiently deals with those. In contrast, Rowhammer flips are not IID: they are generated by a malicious attacker "hammering", and show distinct, strongly correlated access patterns and bit flips. Moreover, ECC corrects bit flips *after* they happen, but it's better to *prevent* them from happening altogether. Hence ECC in DRAM is sub-optimal as mitigation against Rowhammer.

Dedicated Rowhammer mitigations are fine-tuned to the attack: known Rowhammer attacks exhibit highly characteristic row[2] access patterns which are deemed to be easily trackable with a small number of counters, hence little silicon area is lost. If those counters indicate that rows are accessed in a suspicious way, we can prematurely refresh neighbouring rows before a flip happens. This is called TRR (= target row refresh), used, seemingly, in all modern DRAM, and is described in more detail in Section 2. As ambient radioactivity remains a problem, TRR is used in conjunction with ECC. Our learner discovers parameters for both.

*Towards learning the space of Rowhammer mitigations.* Both ECC and TRR have a large parameter space, e.g., block and message lengths for Reed–Solomon codes. Since details of TRR are not public, we have to deal with many additional uncertainties, for example: what does it mean for rows to be accessed in suspicious ways? Which and how many neighbouring rows are considered under attack? When are premature refresh commands issued, and for what rows? Ideally we would like a tool that we can attach to a physical DRAM and it learns the exact Rowhammer mitigation the chip provides with substantial precision. Such a tool does not currently exist, since there are numerous difficulties to overcome, in addition to the large parameter space for ECC and TRR:

– Complex DRAM-access signalling protocols that must be adhered to, but are not of interest to our learning mechanism.
– DRAM signalling runs at nano-second speed, which is hard to reconcile with the speed of software-based learning mechanisms, hence hardware adaptation [14] is required.

We believe that overcoming all of them in one go is currently infeasible. Instead we split them into three largely orthogonal tasks.

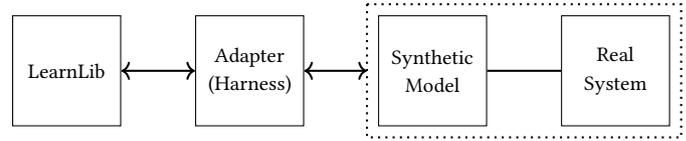

Figure 1: 'Recipe' for using LearnLib against real systems.

– First design learning mechanisms against synthetic models of DRAM, to develop the abstractions and infrastructure that connects the generic learner with the world of DRAM.
– Use the off-the-shelf testing harness [3] to fine-tune the learner so it handles DRAM signalling.
– Finally learn and validate parameters of physical DRAM.

The present work is concerned only with the first task, but the long-term goal is to connect our learner to physical DRAM. We now discuss what learning mechanism to use?

*Active learning.* Despite deep learning's popularity, we use *active learning*, which allows inferring an automaton model of a black-box system by observing its output in response to queries. This is opposed to *passive* learning, such as PAC learning [29], where the training data is provided and no further interactions with the system are possible. There are two main reasons for using active learning. First, unlike neural networks, learned automata are easily *interpretable*. This is important for us, since our's is not a simple supervised learning problem: we have no labelled training data, and we do not know what the exact mitigations are that DRAM is using. Instead, we must refine our learning process in response to what it does and does not learn, and compare that with the assumptions about TRR in the literature. Such an iterative refinement would be difficult if the result of learning was a matrix of floating point numbers. Second, we are motivated by the series of successful applications of active learning to reverse-engineering real-world hardware and software systems [5–9, 30].

We use LearnLib, a state-of-the-art active learning library [11, 16] to automatically infer an automaton model of the behaviour of a DRAM module under Rowhammer attacks. The model will represent memory accesses and their effects, including bit flips and mitigations being applied. The core problem is the large state-space of Rowhammer mitigation mechanisms, so our challenge is to identify which information is relevant and thus needs to be included in the model, and which information can be ignored or abstracted away to make learning more scalable.

Exploring a concrete system $S$ with an active learner, visualised in Figure 1, involves a few generic steps (see [1] for details):

**Translation.** Translate between the language of the active learner and that of $S$. This interaction is usually bidirectional.
**Abstraction.** $S$ is often infinite and also contains information not of interest to the learning process. This means the translation process is also an abstraction (when the information flows from $S$ to the learner) and a corresponding concretisation (in the other direction). Sometimes, the abstraction layer does not talk to $S$, but instead represents $S$ by an additional layer of simplification, the **synthetic model** of $S$.
**Adapter.** Translation, abstraction and concretisation is typically handled by an adaptation layer (aka harness, connector,

---

[2]Modern DRAM do not offer access to individual bits, but rather to only rows containing $2^k$ many bits. The size $k$ is not important for the attack to work, only that bits we access have neighbours! It's the neighbourhood relation on a DRAM that determines which bits are in danger of being flipped.

or mapper), which has two interfaces, one with the learner and another with S.

**Determinisation.** In addition, if S is non-deterministic, as most real-world systems are in some form, we must make the learner's view of S deterministic.

Subsequent sections explain how we instantiate those steps.

*Our contributions.* In summary, our contributions are as follows:
- A simple yet expressive automaton model of DRAM behaviour under Rowhammer attacks.
- The first formal model of TRR, including a new parameter, TRR size, that we believe is key for analysing TRR.
- The first approach based on active learning for automatically determining core parameters of realistic Rowhammer mitigation mechanisms.
- An implementation of our techniques as an open-access tool ALARM. All code is available from [26].
- A detailed set of experimental results.

## 2 MODELLING ROWHAMMER AND ROWHAMMER MITIGATIONS

The target of our learning process will be a synthetic model (see Figure 1) of DRAM with TRR, ECC, and probabilistic bit flips. The learned model, on the other hand, will be much simpler – a deterministic Mealy machine – providing a higher-level of abstraction; in particular, any DRAM state information is abstracted away. The remaining information lets us achieve our goal to read off easily the TRR and ECC parameters of the system learned. We first discuss Rowhammer mitigations and insights that informed the design of our synthetic model, then we provide details of the two aforementioned models.

*Mitigations overview.* DRAM uses both ECC to mitigate non-malicious faults and TRR as Rowhammer-specific mitigation. ECC is well-known, so, for reasons of space, we only detail TRR here.

The key insight enabling mitigations is that Rowhammer attacks have a clear attack pattern: *"Rowhammer: It's like breaking into an apartment by repeatedly slamming a neighbor's door until the vibrations open the door you were after"* [12]. In more detail: in order to flip a bit $b$, we must, in quick succession and between two refreshes, read from or write to bits that are physically close $b$. In DRAM we cannot address individual bits, only rows, so in order to flip a bit in a victim row, we need to read or write physically close "aggressor" rows, between refreshes.

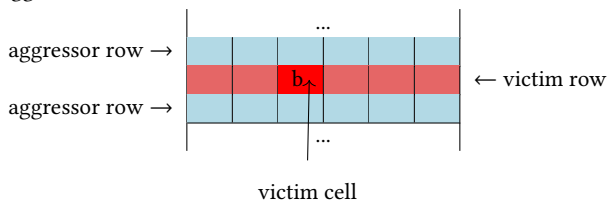

victim cell

This suggests a clear mechanism for detecting and preventing Rowhammer attacks
- Track rows that are accessed in rapid succession between refreshes, called "potential aggressor rows".
- Assume that rows close to potential aggressors are "potential victim rows".
- If the number of accesses to potential aggressor rows exceeds a limit, prematurely refresh potential victim rows.

While the overall principle is clear, there are numerous details that can vary.
- What does it mean for two rows to be neighbours to be close? Relevant is the standard euclidean distance on chip, but should only the immediate neighbour be considered at risk, or, more generally, the closest $n$ rows on both sides? What is the right choice of $n$?
- What exactly is a TRR counter here? We can use precise TRR counters that track row access exactly, or we can use sampling.
- How many rows are being tracked: if we track $k$ rows naively, then a saturation attack that uses more than $k$ aggressors would overwhelm the aggressor tracking capability.
- What kind of temporal discounting to use: if a TRR counter tracks attacks on a victim row $r$ and the attacks move from $r$ to another row, when should the information aggregated for $r$ be forgotten?
- How precise should row tracking information be? For example if row addresses have $s$ bits but TRR counters associate only $t$ bits of the address, with $t < s$, then TRR counters effectively track multiple rows. This has advantages and disadvantages.
- Aggressor centric vs victim centric tracking, meaning that the TRR keeps track of either the potential aggressor rows to find the victim rows, or the potential victim rows to be able to refresh the victims before they get flipped.

There are numerous other parameters, see [13] for a discussion. Unfortunately, DRAM manufactures keep details confidential, and our work is in parts motivated by wanting to provide tooling that gives DRAM customers access to such information.

*Synthetic DRAM model.* In order to develop our active learner in a simplified setting before working with physical DRAM, we have implemented a model of relevant DRAM behaviour in Java. Our model has four main components: Memory, Environment, ECC and TRR. We will now briefly describe each component.

We model memory as a map $\sigma : \text{Row} \to \text{Val}$ where Row is a set of rows (for example addressed by 32 bit unsigned integers, or an infinite set $\{x_n \mid n \in \text{Nat}\}$ in theoretical models) and Val is a set of values that are stored in each row (e.g., individual bits, bytes, *i.e.*, 8 bit unsigned integers, or $\mathbb{Z}$). For simplicity, we assume that each row contains a single value. Memory also provides functionalities for reading from and writing, we omit the straightforward details. Real DRAM distinguish between *internal* ("topological") and externally visible addresses. As this is an orthogonal concern we ignore this distinction. The distance between rows is a core parameter for Rowhammer, both in terms of where flips can happen and how they can be mitigated. In order to model this, we assume that Row itself is a metric space [28] with a notion of distance $\text{dist}(row_1, row_2)$.

In order to orchestrate running a DRAM, and handling Rowhammer mitigations, we need to maintain global information, for example time since the last refresh, the number of accesses to each row since the last refresh, etc. We call this the DRAM's *environment*. For simplicity, and without loss of generality, rather than issuing one refresh command at a time, the environment refreshes all the rows together. This will reset all TRR counters associated with Rowhammer mitigations.

Modelling bit flips is simple: we associate a *flip probability* with read and write operations. Three additional core global parameters in our model that affect flip probability are the *refresh interval*, the *blast radius*, and the *Rowhammer threshold*. The first is the time between two regular refreshes (*i.e.*, not in reaction to a potential Rowhammer attack), the second is the maximum distance between attacker row and victim row that can make bit flips happen. The last is the minimal number of accesses between two refreshes to rows in the blast-radius, before a flip can happen. Hence $row_1$ is a *neighbour* of $row_2$ if

$$row_1 \neq row_2 \quad \text{and} \quad \text{dist}(row_1, row_2) \leq \text{blast radius}.$$

There are other parameters – e.g., time since last refresh, which we omit for brevity, see our code for details, or our forthcoming [27] for a detailed mathematical description of probabilistic models of Rowhammer.

However, the algorithms included in LearnLib only work for deterministic systems under exploration. We determinise our model by being overly conservative: we fix flip probabilities to 1 when the number of accesses (before the next refresh) is above the Rowhammer threshold. Since Rowhammer mitigations must work even in the worst-case (*i.e.*, flips happens as early as possible) to be effective, we believe this to be a reasonable compromise between realism of modelling, and our ability to use existing, well-honed learning tools. The use of sampling to deal with probabilistic behaviour of physical DRAM is future work.

Our ECC component uses the *Reed-Solomon* code for error correction (nothing relies on this choice, we have also implemented *Hamming(8,4)* as an alternative). When ECC corrects a bit flip, our model raises a signal to notify the user that ECC was triggered. This is not far fetched, as in reality ECC being triggered can be detected via a side channel attack [20].

Finally, TRR, the most important component, implements a range of TRR policies. For instance, we have implemented the TRR policies observed for vendors A,B and C in [13].

We note that different ECC and TRR mitigations can be integrated in the model without needing to adapt our tool, thanks to the black-box nature of the learning process (see Section 3).

*Rowhammer machines.* The automata learned by our tool are deterministic finite-state Mealy machines, which are well-understood and widely used models. LearnLib includes efficient learning algorithms for these (e.g., TTT [15]). We introduce a specific class of Mealy machines, which we call *Rowhammer machines*, whose states are abstractions of relevant DRAM state (for example the number rows we wish to model; hereafter we fix a set $n$ memory rows, which we represent as natural numbers $1, 2, \ldots, n$). We explain them in more detail, since they are used in most of this paper.

*Definition 2.1.* A *Rowhammer machine* is a tuple

$$\langle Q, q_0, \text{ACC}, \text{OBS}, \delta \rangle$$

where $Q$ is a finite set of states including a distinguished *flip state* $\perp$, and a distinct initial state $q_0$. We also have finite input and output alphabets ACC and OBS. The input alphabet ACC represents repeated accesses to the $n$ rows: $a \nearrow r \Rightarrow f$ stands for row $r$ being accessed $a$ times, with the intent of flipping $f$ bits. The output alphabet OBS describes the observable effects of those accesses: ✓ if no flips are observed in the neighbouring rows; Flip if the intended number of flips is observed in one of the neighbouring rows; TRR if TRR is triggered; ECC if ECC is applied. Finally, we have the transition function $\delta : Q \times \text{ACC} \to \text{OBS} \times Q$ such that $\delta(q, i) = (\text{Flip}, q')$ implies $q' = \perp$, for all $q, q' \in Q$ and $i \in \text{ACC}$; in other words, observing a Flip always leads to the flip state.

We write $q \xrightarrow{i \,/\, o} q'$ for a transition from $q$ to $q'$ with input $i$ and output $o$, formally: $\delta(q, i) = (o, q')$. The flip state acts as a sink because we are only interested in DRAM behaviour until flips occur.

*Example Rowhammer machines.* We will now give some examples of Rowhammer machines modelling a range of scenarios.

*Example 2.2.* Suppose we consider 2 rows. We want to model a situation in which we access those rows multiples of 100 times with the intent of flipping 1 bit. The Rowhammer threshold for both rows is 120 and no Rowhammer mitigations are in place. The input alphabet is as follows:

$$\text{ACC} = \{100 \nearrow 1 \Rightarrow 1, 100 \nearrow 2 \Rightarrow 1\}$$

The Rowhammer machine modelling our scenario is (loops over the flip state are omitted):

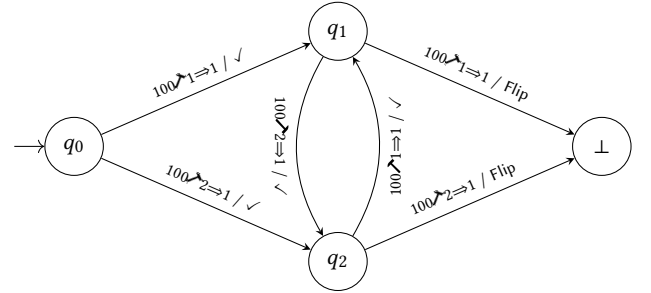

States $q_1$ and $q_2$ are memory states in which the flip has not happened yet; it takes an additional 100 accesses for the flip to occur in a neighbour of row 1. Note that the machine is not able to exactly characterise the Rowhammer threshold, namely 120: it only tells us that a flip is observed between the 100th (exclusive) and the 200th (inclusive) access, *i.e.*, the threshold in the interval $(100, 200]$. The top and bottom paths are symmetrical because we are assuming that the two rows have the same Rowhammer threshold.

The example above assumes knowledge of the Rowhammer threshold. This is not usually the case, and indeed one of our goals is to *learn* it. In order to do so, it is useful to include a range of inputs, with lower numbers of accesses. For instance, the input alphabet ACC = $\{50 \nearrow 1 \Rightarrow 1, 50 \nearrow 2 \Rightarrow 1\}$ will provide a better estimate for the example above, namely (100,150]. Allowing for multiple accesses in one transition makes the model more scalable, however the larger the steps are the less expressive the model is and vice-versa. We will see in Section 4 that this capability brings a trade off between expressiveness and runtime of the learning algorithm.

*Example 2.3.* Under the same scenario as Example 2.2, suppose now that the DRAM supports ECC that can detect and correct 1-bit data corruption. In order to expose it, we may consider an alphabet with a range of target bit flips. For instance:

ACC = $\{100↗1⇒1, 100↗2⇒1, 100↗1⇒2, 100↗2⇒2\}$

Then our model would include the following transitions:

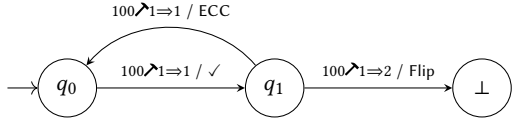

Here ECC is able to correct 1-bit flips, hence the transition back to $q_0$, however 2-bit flips are not successfully corrected, hence the transition to the flip state.

*Example 2.4.* Suppose now that the DRAM supports a simple TRR policy which refreshes a row after 110 accesses. In the automaton of Example 2.2, the TRR refresh of row 1 is represented by an extra transition going from $q_1$ back to $q_0$ (and a similar transition from $q_2$ to $q_0$ for row 2):

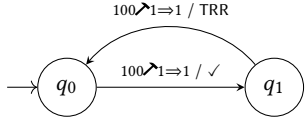

Also, transitions to the flip state are removed, as it is no longer possible to reach the 120 consecutive accesses needed for bit flips to occur. Note that this policy, although appealing, is impractical for real-world DRAM, as it requires one TRR counter per row. Therefore Rowhammer attacks are still possible for realistic TRR policies and we shall see that our tool is able to expose them.

*Example 2.5.* In the previous example TRR led to the initial state. This is not necessarily the case, as some rows may be left in a more vulnerable state in more complex scenarios, and this needs to be recorded in the automaton. For instance, suppose we have 4 rows. We have the following (relevant) transitions:

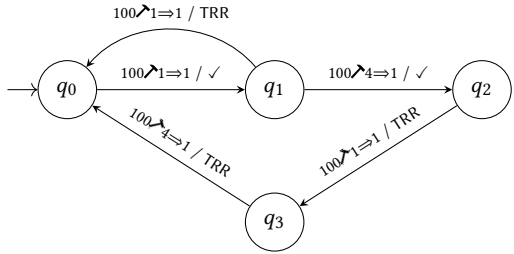

Here we have accessed row 4 in transition from $q_1$ to $q_2$. In this state ($q_2$), accessing row 1 will get TRR triggered, but will no longer take us back to $q_0$, since the impact of the accesses to row 4 have made the neighbouring rows more vulnerable, hence TRR may be triggered again. Indeed, the automaton will move to another state ($q_3$), from which TRR happens.

## 3 ALARM

Our tool ALARM, available at [26], relies on LearnLib's automata learning algorithms to learn a Rowhammer machine from the synthetic DRAM model; it then uses this automaton to infer key mitigation parameters. The high-level architecture, instantiating the approach of Figure 1, is shown in Figure 2. We now describe how the tool works in detail.

*Active learning.* Active learning can be seen as a game where a *learner* interacts with a *teacher* via queries, in the attempt to build an unknown target Mealy machine $M$. The learner can pose two kinds of queries:

- *Membership query:* What is the output of a given input trace $\sigma$? The teacher replies with the output sequence observed when executing $\sigma$ on $M$.
- *Equivalence query:* is a given *hypothesis* Mealy machine $H$ equivalent to $M$? If it is, the algorithm terminates; otherwise, the teacher has to provide a *counterexample*, *i.e.*, a trace distinguishing $H$ and $M$.

The learner will iteratively refine the model by incorporating the query results. When an equivalence query is answered positively, the algorithm terminates, and the learner will output the current hypothesis $H$.

When applying active learning to infer models of real-world systems, the target Mealy machine is an abstraction of the target system's behaviour; membership queries are implemented as tests, and equivalence queries are approximated via membership queries using a conformance checker such as Random Walk [2] and WP-method [17]. The main challenge here is to implement the teacher as an abstraction layer between the learner and the DRAM, capable of translating abstract membership queries as generated by LearnLib to concrete tests and, vice-versa, concrete outputs to abstract ones. For this purpose, we have implemented an intermediate Rowhammer Adapter, which we now describe in detail.

*Rowhammer Adapter.* The target model we want to learn is a Rowhammer machine. For this, LearnLib will generate queries in the form of input sequences from ACC, and is expecting outputs as sequences from OBS. The purpose of the Rowhammer Adapter is two-fold: a) it translates queries from LearnLib to the target (and vice-versa); b) it records information that is relevant for the synthetic DRAM model, but not for the learner.

As an example, in Figure 2 we see that each action within the given query is translated by the Rowhammer Adapter to a sequence of read accesses to the concrete rows $0x10$ and $0x11$, with the intent of flipping 2 bits. The Rowhammer Adapter keeps track of the content of the neighbouring rows (*i.e.*, the rows within blast radius = 1 in this example) and, after the accesses are performed, it will check whether the content of those rows match its internal records. In case they don't, it will return Flip whenever the number of flipped bits matches the input action, namely 2. In all other cases (*i.e.*, the number of flipped bits is either 0 or 1) it will return ✓. In Figure 2, 2 bit flips have been observed in row $0x10$ after the last sequence of accesses, therefore a Flip output is returned.

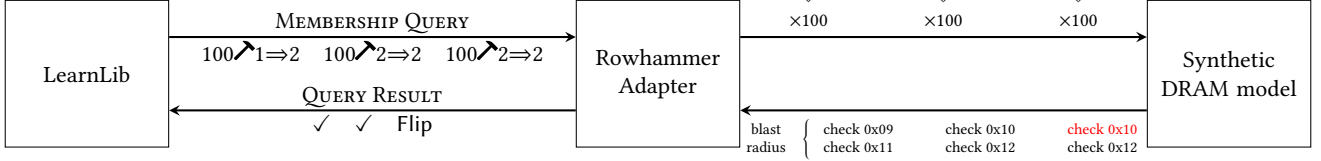

Figure 2: ALARM architecture.

We note that the concrete row addresses and their content are abstracted away by the Rowhammer Adapter. This information is irrelevant for our purposes, and hiding it from the learning algorithm is key for scalability.

Another responsibility of the Rowhammer Adapter is to check whether TRR and ECC have been triggered and to inform the learner via the correct outputs, namely TRR and ECC. In our synthetic DRAM model this is achieved via Java exceptions, in reality the adapter will have to apply suitable heuristics (see, e.g., [13, 20]) to detect the activation of Rowhammer mitigations.

We note that the adapter supports *any* implementation of TRR and ECC. In fact, interactions with the synthetic model are via reads in a purely black-box fashion.

*Inferring mitigation parameters.* Once a Rowhammer machine has been learned, ALARM will use it to infer important features of Rowhammer mitigations. We now detail how this is achieved. We will also define a new concept – *TRR size* – which provides a uniform measure for quantifying the effectiveness of TRR policies. This measure is enabled by the abstraction power of a Rowhammer machine, which only captures the behaviour of those policies, ignoring implementation details.

*Rowhammer threshold.* To extract an estimate of the Rowhammer threshold (see Section 2 for the definition) from a Rowhammer machine, we perform a breadth-first search of the underlying graph, starting from the initial state end ending at the flip state. The shortest path (or any of them) between those states will allow us to estimate the Rowhammer threshold, and the sequence of accesses in that path may be considered as a program capable of inducing bit flips. More precisely, suppose that the shortest path is:

$$q_0 \xrightarrow{a_0 \nearrow r_0 \Rightarrow f_0 \,/\, o_0} \cdots \xrightarrow{a_{n-1} \nearrow r_{n-1} \Rightarrow f_{n-1} \,/\, o_{n-1}} q_n \xrightarrow{a_n \nearrow r_n \Rightarrow f_n \,/\, \text{Flip}} \bot$$

Let $A$ be the number of accesses until state $q_n$, namely $A = \sum_{i=0}^{n-1} a_i$. Then the Rowhammer threshold is in the interval $(A, A + r_n]$. Note that any other paths between those states will give us a sequence (hence a program) capable of triggering bit flips, albeit longer.

*Example 3.1.* Consider the model below, with 1 access per step:

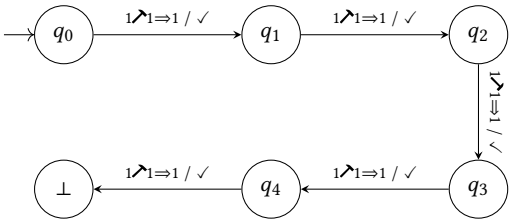

We see that the Rowhammer threshold is given by the path length, namely $(4, 5] = 5$. However, to achieve better scalability and obtain a more compact model, we can increase the number of accesses in the input alphabet, for instance:

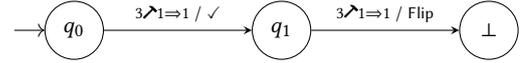

We now have a smaller model, but the estimate is coarser – $(3, 6]$.

*TRR threshold.* The TRR threshold is the minimum number of accesses to a row required for TRR to refresh neighbouring rows that are considered potential victims.

To extract the TRR threshold from the output model, we first look for the shortest cycle in the underlying graph which contains a TRR output. Then we can compute an interval estimating the TRR threshold from that cycle by adding up the accesses before TRR is triggered, similarly to what we did for the Rowhammer threshold. If several such shortest cycles exist, any of them will provide a valid estimate, and a sequence of accesses that can trigger TRR.

*Example 3.2.* Consider the Rowhammer machine below:

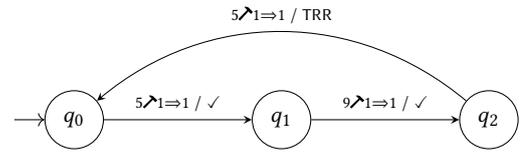

The cycle involving all nodes gives us an estimate of the TRR threshold in $(14, 19]$. We can further refine the estimate by reducing the number of accesses in the input alphabet.

*TRR size.* TRR policies rely on TRR counters to keep track of accesses to rows that may cause a bit flip. The quality of a TRR policy does not solely depend on the number of TRR counters; perhaps more crucially, it depends on the way those TRR counters are used. We define a new concept, *TRR size*, that measures how effectively TRR counters are used.

*Definition 3.3 (TRR Size).* Suppose the DRAM supports TRR. The TRR Size is the minimum number of rows that need to be accessed in order for a bit flip to occur.

To illustrate this concept, we shall now go through three examples. In all three cases we assume that there are two TRR counters. We also assume that TRR threshold and Rowhammer threshold values are close.

*Example 3.4 (Static policy).* Let's assume we have the memory layout in Figure 3a, and the Rowhammer blast radius is 1. The TRR policy has statically assigned each TRR counter to keep track of accesses to *0x10* and *0x11* respectively. In this case, one can hammer *0x14* and cause flips in *0x13* without TRR being able to stop it. The TRR size for this policy is 1 since we were able to cause a bit flip by accessing only 1 row (namely *0x14*).

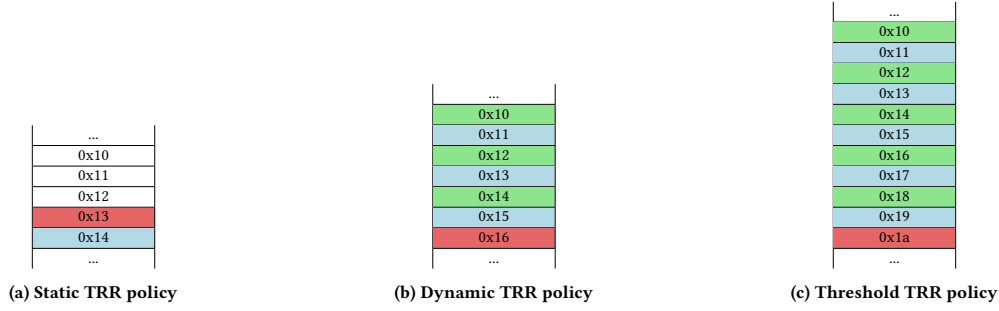

(a) Static TRR policy  (b) Dynamic TRR policy  (c) Threshold TRR policy

Figure 3: Example TRR policies. Grey rows are potential aggressor rows, green ones are potential victims that are refreshed by TRR, and red ones are victim rows.

*Example 3.5 (Dynamic policy).* Consider now the layout in Figure 3b. Now the TRR policy keeps track of accesses to the *first and second* rows that are accessed after a refresh. Suppose we first access *0x11* and then *0x13*, so that both TRR counters are assigned (one to *0x11* and one to *0x13*), then we access *0x15*, and we start over from *0x11*. Here we observe bit flips in *0x16* because TRR is not able to keep track of the accesses to *0x15*. The TRR size for this policy is 3 since we were able to cause a bit flip by accessing three rows, namely *0x11*, *0x13*, and *0x15*.

*Example 3.6 (Threshold policy).* Consider now the memory layout in Figure 3c. The policy for this example keeps track of accesses to the first and second rows that are accessed. When one of the TRR counters reaches the threshold $k$, all the potential victims of the associated row are refreshed and the TRR counters are assigned to new rows. In addition, if a row is not accessed again within a time threshold $t$, its TRR counter is reassigned and the potential victims are refreshed.

Suppose $k = rowhammer\_threshold/2$ and $t$ is equivalent to the time needed to access a row a few times (e.g., 5 times). And we access the grey rows in the figure in a loop going from 0x11 to 0x19. At first the two TRR counters are assigned to *0x11* and *0x13* respectively. Once the TRR counter for *0x11* reaches $k$, its victims are refreshed, and the TRR counter is re-assigned to *0x15*; similarly for *0x13*, with its TRR counter being re-assigned to *0x17* when it reaches $k$. Now, *0x19* has been accessed without TRR tracking it, so *0x1a* is getting hammered and bit flips will eventually occur. In this example the TRR size is 5, because we were able to cause a bit flip by accessing rows *0x11*, *0x13*, *0x15*, *0x17* and *0x19*.

Note that all three examples rely on two TRR counters, but these are used differently: the first policy is the least effective – bit flips may occur due to accessing only one row, *i.e.*, , the TRR size is 1 – and the last one is the most effective – bit flips require accesses to five rows, *i.e.*, , the TRR size is 5. In other words, a TRR policy cannot be evaluated on the basis of the number of TRR counters alone; we need a behavioural model such as our Rowhammer machines to fully appreciate the policy's effectiveness.

Again, the TRR size can be extracted via a visit. Now we are looking for a path between the starting and flip states containing the minimum number of accessed rows; this number is exactly the TRR size. Such a path will give us an efficient – in the sense of the number of rows getting accessed – way to bypass TRR.

*Example 3.7.* Consider the following Rowhammer machine, modelling 4 rows. The DRAM supports a similar TRR policy as Example 3.5, with one TRR counter to count accesses; refresh happens after after 110 accesses. The Rowhammer threshold is 200 (irrelevant state and transitions are not shown):

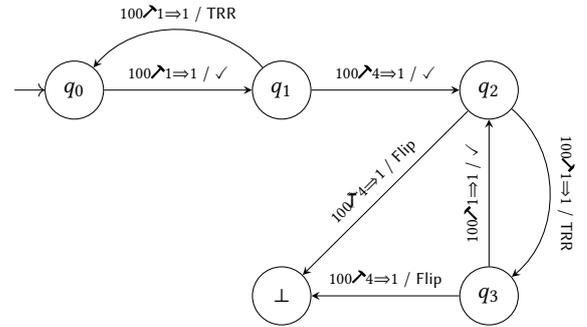

The path mentioning the fewest rows is $q_0 \rightarrow q_1 \rightarrow q_2 \rightarrow \bot$ – namely, it mentions 1 and 4 – therefore the TRR size is 2.

*ECC threshold.* We call *ECC threshold* the maximum number of bits that ECC can correct. To extract the ECC threshold from the model, we take the maximum $f$ over all transitions of the form

$$q_i \xrightarrow{a \nearrow r \Rightarrow f \,/\, \text{ECC}} q_j \ .$$

## 4 EXPERIMENTS AND MEASUREMENTS

In this section we measure the runtime of our tool against different configurations of our synthetic memory. The TRR policy used in the experiments is a victim centric policy that assigns a TRR counter on a first-come-first-serve way, and we use Reed-Solomon ECC with the ability to correct 4 bits. Models can be successfully learned in most cases, however we shall see that LearnLib parameters affect runtime and accuracy of the output model. In the experiments, we focused on one parameter at a time and fixed the rest on defaults values shown in the table on the next page.

*Memory size.* Memory size has a direct impact on the runtime, *i.e.*, the larger the memory size the longer the runtime. Particularly, with a larger memory size, the model not only has many distinct memory access patterns, but also more TRR, ECC and Flip events. However, as Figure 4a suggests, the runtime after eight rows is unstable. This is because the Random Walk algorithm with 100 steps is unable to explore the target system in sufficient depth to find a counterexample that the learner can use to further refine the model. This is, in general, a caveat of practical active learning: parameters of the conformance checker may affect the accuracy of the resulting model.

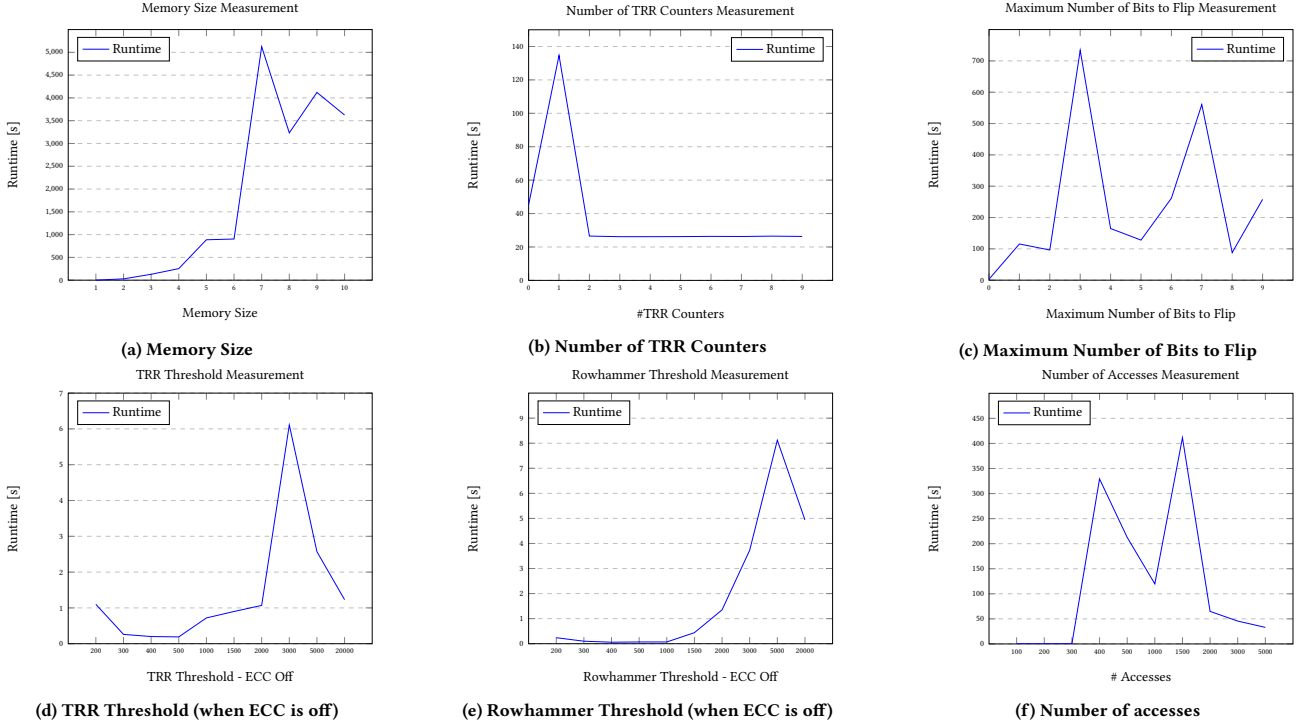

Figure 4: The Impact of Different Parameters on the Runtime

| Parameter | Default Value |
| --- | --- |
| Learning Algorithm | TTT |
| Equivalence Oracle | Random Walk |
| Memory Size | 3 |
| TRR Counters | 1 |
| Maximum Number of Bits to Flip | 6 |
| Number of Accesses in Each Transition | 1300 |
| TRR Threshold | 2500 |
| Rowhammer Threshold | 3000 |
| Refresh Interval | 6500 |
| Blast Radius | 1 |
| TRR Radius | 1 |
| Maximum Steps for the Random Walk Algorithm | 100 |

*TRR counters.* TRR counters have a direct impact on the TRR, as explained in Section 3. Figure 4b illustrates the relation between TRR counters and the runtime. We observe that the runtime stays the same when the number of TRR counters is greater than 2. In fact, for the involved TRR policy, two TRR counters are enough to prevent any bit flips; adding more TRR counters will not add anything to the model.

*Maximum number of bits to flip.* We investigate the impact of bit flips on the runtime by changing the input alphabet. The results are shown in Figure 4c. The x axis in Figure 4c is the maximum number (inclusive) of bit flips in the memory; for example, 2 means that flips of 0, 1 and 2 may occur if accesses to a row reach the Rowhammer threshold before the next scheduled memory refresh. We see that the runtime goes up until 3 bits, and becomes unstable afterwards, because once the maximum number of bit flips is more than 4, the state-space explodes and cannot be fully explored by the Random Walk conformance checker.

*TRR threshold.* In these experiments, we turned off the ECC mechanism for less noise. The result is in Figure 4d. When TRR threshold is smaller than 2000, active learning outputs a model almost immediately. This is because TRR protects the memory from the majority of bit flips and the resulting model is significantly smaller. The runtime peaks at 3000 – which is equal to the Rowhammer threshold – due to the learned model being large. Yet once we move beyond 3000, the runtime decreases since bit flips occur and several transitions now end up in the sink state, which results in a smaller model.

*Rowhammer threshold.* ECC is off for this experiment to avoid noise. The results are presented in Figure 4e. The Rowhammer threshold is influenced by the TRR threshold and refresh interval. When the Rowhammer threshold is smaller than the default TRR threshold at 2500 accesses, the runtime is low, because the sink state is quickly reached, hence the model is smaller. When the Rowhammer threshold is above the refresh interval at 6500 accesses, we observe the runtime dropping because the flip state can no longer be reached, and the model has fewer states and transitions. Therefore, as expected, we observe that the model is at its highest complexity when the Rowhammer Threshold resides between TRR Threshold and refresh interval.

*Number of accesses.* Varying the number of accesses, as was explained in Definition 2.1, increases scalability of the model. The larger it is the smaller and the less precise the model is. In Figure 4f, we observe that in the beginning the runtime is low. This is because, for smaller access numbers, models generated by the learner are bigger, and the Random Walk algorithm is unable to cope with them, therefore a smaller, incomplete model is generated. However, as the figure shows, after 1500 accesses the model is simpler but more accurate, and the runtime decreases.

## 5 CONCLUSION

The most up-to-date publicly available information on Rowhammer mitigations, in particular TRR, is [13, 19], which has strongly informed our modelling. Both papers involve a lot of 'manual' labour, and use FPGAs to observe concrete DRAM hardware. Neither uses systematic, ML-based techniques to explore the Rowhammer mitigation space automatically. The present work is complementary to [13, 19] and aims to automate such exploration. Vendors test reliability of DRAM chips using standardised test, see [25, 32]. Knowledge of the physical memory layout is required to run such tests, so this is often not possible for DRAM customers, and [24] show that these tests do not reliably detect all cells that might be vulnerable to Rowhammering. The active learning community has done impressive work on reverse engineering systems [4–6, 8, 9, 30]. Like us, they follow, broadly, the "recipe" from Figure 1, and most have to deal with issues such as non-deterministic behaviour, for instance unpredictable packet arrival times in networking protocols such as OpenSSL [4], SSH [9], DTLS [8] and QUIC [6]. The most striking application to hardware is [30], where cache-replacement policies are inferred from hardware caches.

As future work we mention that it is interesting to make our model more sophisticated, so it can explore, with more accuracy, further details of Rowhammer mitigations. We expect that more details will be forced on us when we attach our learner to physical DRAM.

## ACKNOWLEDGMENTS

The authors would like to thank the anonymous reviewers for their comments. This research was partially supported by EPSRC (EP/S028641/1).